\documentclass[oneside,british]{article}
\usepackage[latin1]{inputenc}
\usepackage{amssymb}

\makeatletter

\providecommand{\LyX}{L\kern-.1667em\lower.25em\hbox{Y}\kern-.125emX\@}

\usepackage{amsmath}
\usepackage{amssymb}

\usepackage{latexsym}

\usepackage[2cell,v2]{xy}
\UseTwocells
\UseHalfTwocells

\def\VCat{{\cal V}\!\!-\!Cat}  
\def\WCat{{\cal W}\!\!-\!Cat}
\def\TwoCat{2\!\!-\!Cat}

\def\Con1VCat{Conv( 1, {\cal V} )\!\!-\!Cat }

\def\pf{{\it Proof:} }
\def \QED{\hfill\hbox{\hskip 4pt
                \vrule width 5pt height 6pt depth 1.5pt}}
\def \epf{\QED\\}

\newcommand{\A}{\mathbb{A}}
\newcommand{\B}{\mathbb{B}}
\newcommand{\Cd}{\mathbb{Cd}}
\newcommand{\Db}{\mathbb{D}}

\newcommand{\N}{\mathbb{N}}
\newcommand{\Od}{\mathbb{Od}}

\newcommand{\M}{\mathbb{M}}

\newcommand{\C}{\mathbb{C}}

\newcommand{\T}{\mathbb{T}}
\newcommand{\X}{\mathbb{X}}

\newcommand{\V}{\mathcal{V}}
\newcommand{\W}{\mathcal{W}}

\newcommand{\free}{\mathcal{F}}

\newcommand{\one}{id}

\newcommand{\deq}{\stackrel{def}{=}}
\newcommand{\dee}{\stackrel{def}{\Longleftrightarrow}}

\newcommand{\cat}{\emph{\textbf{Cat}}}

\newcommand{\ab}{\emph{\textbf{Ab}}}

\newcommand{\rel}{\emph{\textbf{Rel}}}

\newcommand{\set}{\emph{\textbf{Set}}}
\newcommand{\sset}{\emph{\textbf{sSet}}}

\newcommand{\spa}{\emph{\textbf{Span}}}
\newcommand{\spn}[1]{\spa \left ( #1 \right )}

\newcommand{\ob}[1]{Obj\left ( #1 \right )}

\newcommand{\ka}{K}
\newcommand{\slam}{//}
 \usepackage{verbatim}
 \newtheorem{defn}{Definition}[section]
 \newenvironment{lyxcode}
   {\begin{list}{}{
     \setlength{\rightmargin}{\leftmargin}
     \raggedright
     \setlength{\itemsep}{0pt}
     \setlength{\parsep}{0pt}
     \normalfont\ttfamily}
    \item[]}
   {\end{list}}
 \newtheorem{prop}{Proposition}[section] 
 \newtheorem{thm}{Theorem}[section]
\newtheorem{lemma}{Lemma}[section]

\AtBeginDocument{
  
}

\usepackage{babel}
\makeatother
\begin{document}
\date{}

\title{Bisimulation of enrichments}

\author{Vincent Schmitt and Krzysztof Worytkiewicz}

\maketitle
\begin{abstract}
In this paper we show that classical notions from 
automata theory such as simulation and bisimulation 
can be lifted to the context of enriched categories. 
The usual properties of bisimulation
are nearly all preserved in this new context.
The class of enriched functors that correspond to functionnal
bisimulations surjective on objects is investigated and appears
to be ``very close'' to be open in the sense of 
Joyal and Moerdijk \cite{JoMo94}. 
Seeing the change of base techniques as a convenient 
means to define process refinement/abstractions we 
give sufficient conditions for the change of base categories
to preserve bisimularity.
We apply these concepts to Betti's generalized automata, 
categorical transition systems, and other exotic categories.
\end{abstract}

\section{Introduction}

In \cite{Law73} it is shown that enrichments over a particular monoidal
closed category may capture the notion of generalized metric spaces.
Since these fundamental works, various mathematical objects have been
successully coded as enrichments. The long list includes sheaves \cite{Wal81},
fibrations\cite{BCSW83} and more recently again metric spaces and
also quasi-uniform spaces \cite{Sch03} \cite{Sch04}. Betti introduced
generalized automata qua enrichments in unpublished notes. Results
regarding their minimal realization occur in \cite{KKR83}. In \cite{KaLa99}
occured the idea that $\V $-categories may represent abstract machines 
and the base $\V $ is seen a well structured set of computations.
In this framework, refining the universe of computation becomes a
change of base. This is the point of view that we develop in the present
work.

We apply change-of-base techniques to define``good'' notions
of abstraction/refinement. Seeing $\V $-categories as processes 
performing their computations in $\V $, we claim that any reasonable 
change of base from $\V $ to $\W $ should preserve usual process equivalences.
First we show that the notion of bisimulation
generalizes to enrichments. Though the status of bismimulation
in enriched categorical term remains unclear, 
it is pretty simple and behaves astonishingly well.
To sum up all classical properties of the bisimulation 
of automata lift to enrichments. 
This fact yields a clear notion
of ``good'' changes  of base: the ones preserving bismularity. 
We investigate the conditions under which this happens and derive a 
sufficient one. Our work relies on recent results \cite{KLSS02}, 
\cite{LaSc02} where changes of base are identified as two-sided 
enrichments. The latter are akin to geometric morphisms between 
categories of sheaves. 

Eventually we treat a more elaborate example than Betti's automata
namely the categorical transition systems as an illustration of 
our categorical machinery.

\section{\label{sec:Enr}Enrichments over Bicategories}

Enrichments over bicategories generalize the classical enriched category
theory over monoidal base categories (ECT). The latter being just one-object
bicategories, one could argue that both theories should formally be
the same and it is true to a certain extent. On the other hand, even
classical ECT becomes less so when the base category is \emph{not}
symmetric monoidal. In such a case, the best one can hope for in order
to have a well-behaved theory is \emph{biclosedness.} It is a rather
unusual situation in mathematical practice which deals with enrichments
over $\ab $ or over $\sset $, yet it seems to be standard in Computer
Science where computational paths are generally irreversible. It turns
out that ECT over non-symmetric monoidal categories is conveniently
studied as a special case of ECT over bicategories.
In this paper we shall use enrichments over \emph{locally (pre)ordered}
bicategories. They are in fact very simple
2-categories and their simplicity offers the pleasant fringe 
benefit that we can dispense
with checking coherence conditions. 
In this chapter we shall present briefly elements of the theory
and illustrates the relevance of these constructions with
Betti's automata and other derivatives.\\

In what follows we shall denote by $\otimes $ the horizontal composition
in a bicategory, written in the diagrammatic order.

\begin{defn}
\label{defenrlp} Let $\V $ be \emph{locally preordered} bicategory.
An \emph{enrichment} $\A $ over ${\mathcal{V}}$, also called a \emph{${\mathcal{V}}$-category},
is a set $Obj(\A )$ along with mappings \[
(-)_{+}:Obj(\A )\rightarrow Obj({\mathcal{V}})\]
 and \[
\A (-,-):Obj(\A )\times Obj(\A )\rightarrow Ar({\mathcal{V}})\]
 such that 
\begin{enumerate}
\item $\A (a,b):a_{+}\rightarrow b_{+}$ for all $a,b\in \ob {\A }$;
\item $id_{a_{+}}\leq \A (a,a)$ for all $a\in \ob {\A }$;
\item $A(a,b)\otimes A(b,c)\leq A(a,c)$ for all $a,b,c\in \ob {\A }$.
\end{enumerate}
$\A $'s \emph{fiber} $\A _{x}$ \emph{over $x$} is the set $\{a\in \ob {\A }\mid a_{+}=x\}$
\end{defn}
Given a $\V $-category $\A $, notice that\[
\ob {\A }=\coprod _{x\in \V }\A _{x}.\]

\begin{defn}
Let $\A $ and $\B$ be ${\mathcal{V}}$-categories. A \emph{${\mathcal{V}}$-functor}
$F:\, \A \rightarrow \B $ is a map $f:Obj(\A )\rightarrow Obj(\B )$
such that:
\begin{enumerate}
\item ${(-)}_{+}^{\A }={(-)}_{+}^{\B }\, \circ \, f$;
\item $\A (a,b)\leq \B (fa,fb)$ for all $a,b\in \ob {\A }$.
\end{enumerate}
Given ${\mathcal{V}}$-functors $F,G:\, \A \rightarrow \B $, a \emph{${\mathcal{V}}$-natural
transformation} from $F$ to $G$ is given by the family \[
\left\{ id_{a_{+}}\leq B(fa,ga)\right\} _{a\in \ob {\A }}\]
 of 2-cells in $\V $. 
\end{defn}
Observe that there can be at most one \emph{${\mathcal{V}}$-natural
transformation} from $F$ to $G$. 

\begin{defn}
A bicategory $\mathcal{V}$ is \emph{biclosed} if for any arrow $f:u\rightarrow v$
and any object $w$ of ${\mathcal{V}}$, the functors $- \otimes f:{\mathcal{V}}(w,u)\rightarrow {\mathcal{V}}(w,v)$
and $f \otimes -:{\mathcal{V}}(v,w)\rightarrow {\mathcal{V}}(u,w)$
are left adjoints. (This is exactly to say that $\V$ has all 
{\em right Kan extensions} and {\em right liftings} according to the 
terminology of \cite{DaSt97}). 
\end{defn}
In this work, we shall mostly be concerned with \emph{quantaloids}. 

\begin{defn}
A \emph{quantaloid} is a small, biclosed, locally cocomplete 
and locally ordered bicategory.
\end{defn}
A familiar example of a quantaloid is
$\rel $, the locally-ordered bicategory with
\begin{itemize}
\item objects: (small) sets;
\item morphisms: $\rel (X,Y)$ is the set of the binary relations from $X$
to $Y$ ordered by inclusion;
\item composition: the usual relational composition.
\end{itemize}
\begin{prop}
Let $\mathcal{V}$ be a quantaloid. ${\mathcal{V}}$-categories, ${\mathcal{V}}$-functors
and ${\mathcal{V}}$-natural transformations form a locally preordered
2-category denoted $\VCat $. 
\end{prop}
Given a category $\C $, one can build a quantaloid $B(\C )$ as follows.
$B(\C )$ has the same objects as $\C $. Given $c,c'\in \C $,
the partial order $B(\C )(c,c')$ is the powerset of $\C (c,c')$ ordered
by inclusion. The horizontal composition \[
B(\C )(c,c')\otimes B(\C )(c',c'')\rightarrow B(\C )(c,c'')\]
 is pointwise and $id_{c}^{B\left(\C \right)}=\left\{ id_{c}^{\C }\right\} $.
This bicategory admits right Kan extensions and right liftings.\\ 

Enrichments on $B(\C )$ are lax functors from $\C $ to $\rel $.
Suppose $\A \in B\left(\C \right)\textrm{-}\cat $. There is a lax
functor\[
\begin{array}{rlcl}
 F\left(\C \right): & \C  & \rightarrow  & Rel\\
  & c & \mapsto  & \A _{c}\\
  & c\xrightarrow{{f}}c' & \mapsto  & \left\{ \left(a,a'\right)\in \A _{c}\times \A _{c'}\, \mid \, f\in \A \left(a,a'\right)\right\} \end{array}
\]
The construction reverses.\\

In order to emphasize the interpretation of $B\left(\C \right)$-categories
as computational devices, we use the automata-theoretic notation $a\xrightarrow{{m}}b$
for the arrow $m:a_{+}\rightarrow b_{+}\in \A (a,b)$.

A $B(\C )$-functor $F:\A \rightarrow \B $ is just a map $F$ from
the states of $\A $ to those of $\B $, such that

\begin{enumerate}
\item $F\left(\A _{c}\right)\subseteq \B _{c}$ for all $c\in \C $;
\item $a\xrightarrow{{m}}a'\Rightarrow F\left(a\right)\xrightarrow{{m}}F\left(a'\right)$
for all $a,a'\in \ob {\A }$.
\end{enumerate}
In other words, $F$ is a \emph{functional simulation}. A $B(\C )$-natural
transformation $F\Rightarrow G:\A \rightarrow \B $ is given by the
family of arrows

\[
\left\{ F\left(a\right)\xrightarrow{{id_{a^{+}}}}G(a)\right\} _{a\in \ob {\A }}.\]

Recall that monoidal categories may be seen as one-object bicategories
and monoidal functors correspond to lax functor between those.
In the same way, \emph{quantales} are one-object quantaloids. Seen as 
categories (more precisely as \emph{monoidal partial orders}),
quantales are complete and cocomplete.\\

Given a monoid $\M$, the \emph{quantale of $\M$-languages}
$C(\M)$ (seen as a monoidal category) has objects the subsets of 
$M$ ordered by inclusion, its unit is the subset $\{id_{\M}\}$. 
The tensor of $C(\M)$
is the pointwise composition \[
L\otimes L'=\{l\cdot l'\mid l\in L,l'\in L'\}\] for $L,L'\in \wp (\M)$.
This quantale is generally \emph{not} symmetric but always bi-closed,
by construction. $C(\M)$-enrichments were called \emph{generalized
automata} by Betti in his unpublished notes.\\ 

An extra motivation for considering enrichments over quantaloids
rather than just on quantales comes from the following observation
regarding slice categories.\\

Starting from a quantaloid $\V $ and a $\V $-category $\A $, the
quantaloid $\V (\A )$ is as follows. Its objects are those of $\A $.
For any $a,b\in \ob {\A }$, $\V (\A )(a,b)$ is the preorder 
$\V (a_{+},b_{+})\downarrow \A (a,b)$.
The composition of $a\xrightarrow{{f}}b\xrightarrow{{g}}c$ in $\V (A)$
is the arrow of $\V $ $f\otimes g\leq A(a,b)\otimes A(b,c)\leq A(a,c)$.
The identity in $a$ is the 2-cell $id_a +\leq \A (a,a):a_{+}\rightarrow a_{+}$.

\begin{prop}\label{comma}
Given a quantaloid $\V$, a $\V $-category $A$, let
$\VCat \downarrow \A $ denote slice 2-category over $A$, then 
there is a 2-isomorphism\[
\V (\A )\textrm{-Cat}\cong \VCat \downarrow \A \]
\end{prop}
Actually the latter isomorphim is natural 
in $\A$ in the following sense. Any $\V$-functor $F:\A \rightarrow \B$
defines a 2-functor $\V(F): \V(\A) \rightarrow \V(\B)$,
sending objects $a$ to $Fa$ and with components the embeddings
$$\V(F)_{a,a'}\; : \; \V(a_+,a'_+) \downarrow \A(a,a')
\rightarrow \V(a_+, a'_+) \downarrow \B(Fa,Fa').$$
If there is a $\V$-natural $F \Rightarrow G: \A \rightarrow \B$
which consists in our specific context in a collection of 
inequations $$id_{a_+} \leq \B(Fa,Ga): a_+ \rightarrow a_+$$ for 
all objects $a \in \A$, then it is also a collection of 
arrows in $\V(\B)$ from $\V(F)(a)$ to $\V(G)(a)$.
This collection defines then a 2-natural transformation
$\V(F) \Rightarrow \V(G)$ as for all $a,b \in \A$,
$- \circ id_a : \V \downarrow \B(Ga,Gb) \rightarrow \V \downarrow
\B(Fa,Gb)$
and $id_b \circ -: \V \downarrow \B(Fa,Fb) \rightarrow \V \downarrow 
\B(Fa,Gb)$ are the obvious embeddings. 
Now one may check that the assignements $\V(-)$ above define 
indeed a 2-functor $\VCat \rightarrow \TwoCat$. 
We shall come back later on the above isomorphism when 
we treat the change of base bicategory.

Computationally, such slices can be seen as computational devices 
with \emph{interfaces}, i.e. \emph{processes}. Given a process 
$\left(\begin{array}{c}
 \X \\
 \downarrow \\
 \A \end{array}
\right)$, we refer to $\X $ as its \emph{implementation} and to $\A $ as
its interface. Put differently, the base (of the slice) $\A $ represents
the part of $\A $ which is \emph{observable.}

\section{Bisimulation}

With Betti's automata in mind, we introduce now a quantaloid-enriched 
version of the well-known simulation and bisimulation for
automata.
These notions extend well to enrichments
and the main results of the theory still hold
in this more gerenal setting.\\
 
In this section $\V$ will denote a quantaloid.\\

Consider $\V$-enrichments $\A$ and $\B$. Let us first
consider a relation $R$ from $\A$ to $\B$, i.e.
$R \subseteq Obj(\A) \times Obj(\B)$, such that 
if $(a,b) \in R$ then $a_{+} = b_{+}$.
$R$ is a {\em simulation} from $\A$ to $\B$ 
if for all $(a,b)\in R$,
$$\forall a' \in \A,\;\; \A(a,a') \leq \bigvee_{(a',b') \in R} \B(b,b')$$
$R$ is a {\em bisimulation} if and only if both 
$R$ and $R^{-1}$ are simulations.\\

For any $\V$-enrichments $\A$ and $\B$, any union 
of simulations (respectively bisimulations)
from $\A$ to $\B$ is a simulation (respectively a bisimulation),
therefore there exist a larger simulation and a larger 
bisimulation from $\A$ to $\B$.
For $a \in \A$ and $b \in \B$ we say that {\em $b$ simulates $a$},
(respectively {\em $a$ bisimulates $b$}) 
if there is a simulation from $\A$ to $\B$
(respectively a bisimulation) $R$ 
with $(a,b) \in R$.\\

With the notation above, let $i_a: \B_a \rightarrow \B$ denotes 
the full subcategory of $\B$ with objects those $b$ with $(a,b) \in
R$.  Then to say that $R$ is 
is a simulation is to say that for any $(a,b) \in R$,
$A(a,-)$ is less than the colimit of $\B(b,-) \circ i_a : \B_a
\rightarrow \V$.\\

Consider a map $f: Obj(\B) \rightarrow Obj(\A)$ such that for all $a
\in Obj(\A), (f(a))_{+} = a_{+}$. It is
a particular relation, say $R$, from $\A$ to $\B$ and as such:\\
- $R$ is a simulation from $\A$ to $\B$ if and only if
$$(1)\;\forall b \in \B, a' \in \A,\;\; 
\A(f(b),a') \leq \bigvee_{ b' \mid f(b') = a'} \B(b,b')$$
- $R^{-1}$ is a simulation if and only if
 $$(2) \forall b,b' \in \B,\;\; \B(b,b') \leq \A(f(b),f(b')).$$
Condition $(2)$ is just that $f$ defines a $\V$-functor
$\B \rightarrow \A$. When $(2)$ holds, $(1)$ is equivalent then to
$$(1')\; \forall b,a' \in \A,\;\; \A(f(b),a') = \bigvee_{b' \mid f(b') = a'}
\B(b,b').$$ So that a functionnal bisimulation amounts 
to a $\V$-functor satisfying $(1')$.\\
 
With the notation above, let $i_a: \B_a \rightarrow \B$ denotes the fiber 
of $f$ over $a$, that is the full subcategory of $\B$ with 
objects those $b$ with $f(b) = a$. Then to say that $f$ satisfies
$(1')$ is to say that the representable module $\A(f(b),-)$
is given by the colimit 
$$A(f(b),a') = colim ( \B(b,-) \circ i_a' : \B_a \rightarrow \V ).$$\\

\noindent We shall say that:
\begin{itemize}
\item {\em $\B$ simulates $\A$} when there exists 
a simulation from $\A$ to $\B$ such that 
\begin{center}
for all $a \in \A$, there exists $b \in \B, (a,b) \in R.$
\end{center}
\item {\em $\A$ and $\B$ are bisimilar} when there
exists a bisimulation $R$ from $A$ to $B$ such that
\begin{tabbing}
- for all $a \in \A$, there exists $b \in \B$ such that $(a,b) \in R$ and\\
- for all $b \in \B$, there exists $a \in \A$ such that $(a,b) \in R.$
\end{tabbing}
\end{itemize}

In particular if $f: \A \rightarrow \B$ is a functional
bisimulation that is surjective on objects then
$\A$ and $\B$ are bisimilar.\\ 

\noindent\textbf{Example 1} [Automata!]
Obviously our notions of simulation, bisimulation extends 
the classical ones for automata. For an alphabet $\Sigma$,
a relation of simulation, respectively bisimulation,
between $\Sigma$-automata $\A$ and $\B$
is exactly a simulation, respectively a bisimulation
between the $C(\Sigma ^{*})$-enrichments $\A$ and $\B$. 
Also morphism of $\Sigma$-automata
$f:A\rightarrow \B$ is a functionnal bisimulation
if and only if
the corresponding $C(\Sigma ^{*})$-functor $f$ 
is.
Remember that these maps were called open 
in \cite{JoyalA:bisom} as their whole class satisfies axioms
$(A1)-(A5)$ for open maps defined by Joyal and Moerdijk.
The maps of the above kind 
are of particular importance in the classical theory of 
bisimulation for automata. As shown further it is also the 
case for the bisimulation of enrichments.\\

\noindent\textbf{Example 2} [Betti's generalized automata]
Things are analogous for Betti's automata.
Let $\M$ be a monoid. Recall from section \ref{sec:Enr} that
a generalized automaton is an enrichment over the quantale
of languages $C(\M)$. A morphism of generalized
automata $f: \A \rightarrow \B$ is 
just $C (\M)$-functor and it defines a surjective bisimulation 
if and
only if for all $a\in \A$, for all $b'\in \B$, for all $m\in \M$,
\begin{tabbing}
If $\xymatrix{ f(a)  \ar[r]^{m} &  b' }$ then there
exists $a' \in \A$ such that $f(a') = b'$ and
$\xymatrix{ a \ar[r]^{m} &  a' }$
\end{tabbing}
One may also easely rephrase in automata theoretic terms,
our notions of simulation and bisimulations for 
$C(\M)$-enrichments.\\

\noindent\textbf{Examples 3,4} [preorders,metric spaces]
In the context of pre-orders (\cite{Law73})
a simulation relation from $\A$ to $\B$ is a  
relation $r \subseteq \A \times \B$ such that 
if $(a,b) \in R$ then for all $a' \geq a \in \A$, 
there exists a  $b' \geq b$ such that $(a',b') \in R$.
In the context of generalized metric
spaces, a simulation relation of $\A$ by $\B$ is a relation 
$r \subseteq \A \times \B$ such that if $(a,b) \in R$ then
for all $a' \in \A$, for all $\epsilon > 0$ there exists a 
$b' \in \B$ such that $(a',b') \in R$ and 
$\A(a,a') \leq \B(b,b') + \epsilon$.
We leave to the reader to decode what bisimulations
mean in those contexts.\\

\noindent\textbf{Example 5} [Simulation/bisimulation over $\A$]
Via the correspondence $ \VCat(\A) \cong  \VCat \downarrow \A$,
simulation and bisimulation relations in $\VCat(A)$ occurs as 
simulations/bisimulations
{\em over $\A$}. That is to say: a simulation over $\A$ of the arrow $f: \B
\rightarrow \A$ by the arrow 
$g: \C \rightarrow \A$ is a simulation $R$ of $\B$ by $\C$ such that
if $(b,c) \in R$ then $f(c) = g(b)$.
Again we leave to the reader to define the bisimulation over $\A$.\\

A few immediate remarks are in order.
It is straightforward to check that simulations/bismimulations 
compose.
Also given a $\V$-enrichment $\A$, the diagonal 
$\Delta_{\A}$ on $Obj(\A)$ is a bisimulation
on $A$ (i.e. from $\A$ to $\A$) and also an equivalence.
So that
\begin{prop}Enrichments and simulation relations ordered 
by inclusion form a locally preordered 2-category.
\end{prop}
Also for any enrichments $\A$, $\B$ and $\C$ if $\B$ simulates
$\A$ and $\C$ simulates $\B$ then $\C$ simulates $\A$ so that 
\begin{prop}
The relation of simularity is a preorder on $\VCat$. 
\end{prop}
Along the same line, bisimulation relations do compose
and as the inverse relation of a bisimulation is a bisimulation,
\begin{prop}
The bisimularity relation is an equivalence on $\VCat$.
\end{prop}

For any enrichement $\A$, the set of bisimulations
{\em on} $\A$ - i.e. from $\A$ to $\A$ contains the diagonal, 
and is closed under composition, inverse and unions. So if $R$ 
is a bisimulation then the equivalence 
$\bar{R}$ generated by $R$ is again a bisimulation.
Bisimulation relations on $\A$ ordered by inclusion
form a complete lattice $Bisim(A)$, 
Bisimulation equivalences on $A$ ordered by inclusion
also form a complete lattice $EqBis(\A)$ and the
map $R \mapsto \bar{R}$ is an upper closure operation
(i.e. for $Bisim(\A)$ and $EqBis(\A)$ seen as
categories, one has a reflection situation:
the inclusion $EqBis(\A) \rightarrow Bisim(\A)$
has as a left adjoint sending any $R$ to $\bar{R}$).\\

Consider now a bisimulation equivalence $\sim$ on $\A$.
It actually defines a ``congruence'' on $\A$ in the following 
sense.
\begin{prop}
The quotient set $Obj(\A) / \sim$ admits 
a $\V$-categorical structure $\tilde{\A}$ and the 
quotient map $Obj(\A) \rightarrow Obj(\A) / \sim$
defines a $\V$-functor 
$\A \rightarrow \tilde{\A}$.
\end{prop} 
If $[a]$ denotes the equivalence class of $\A$ under 
$\sim$, let for any $a,b \in \A$,
$$\bar{\A}(a,b) = \bigvee_{b' \mid b' \sim b} \A(a,b').$$
Then it is immediate that if $b \sim b'$ then for all $a \in \A$,
$\bar{\A}(a,b) = \bar{\A}(a,b')$.
It also happens that if $a \sim a'$ then 
$\bar{\A}(a,-) = \bar{\A}(a',-)$.
To see this let us suppose $a \sim a'$.
So that, since $a'$ simulates $a$,
$$\forall b,\;\; \A(a,b) \leq \bigvee_{b' \sim b} \A(a',b')$$ 
So for any $b$,
\begin{tabbing}
$\bar{\A}(a,b)$\=$=$\=$\bigvee_{b' \sim b} \A(a,b')$\\
\>$\leq$\> $\bigvee_{b' \sim b}  \bigvee_{b'' \sim b'} \A(a',b'')$\\
\>$=$ \>$\bigvee_{b'' \sim b} \A(a',b'')$\\
\>$=$ \>$\bar{\A}(a',b)$.
\end{tabbing}
Therefore it makes sense to define for $a, b \in \A$,
$\tilde{\A}([a],[b])$ as $\bar{\A}(a',b')$ for whatever
$a' \in [a]$ and $b' \in [b]$.\\

Now for any $a,b,c \in \A$,
\begin{tabbing}
$\tilde{\A}([a],[b]) \otimes \tilde{\A}([b],[c])$
\=$=$
\=$( \bigvee_{b' \sim b}\A(a,b') ) \otimes \tilde{\A}([b],[c])$\\
\>$=$  
\>$\bigvee_{b' \sim b}(\; \A(a,b') \otimes \tilde{\A}([b],[c]) \;)$\\
\>$=$
\>$\bigvee_{b' \sim b}(\; \A(a,b') \otimes \bigvee_{c' \sim c} \A(b',c') \;)$\\
\>$=$
\>$\bigvee_{b' \sim b}\bigvee_{c' \sim c}
(\; \A(a,b') \otimes \A(b',c') \;)$\\
\>$\leq$
\>$ \bigvee_{c' \sim c} \A(a,c') $\\
\>$=$
\>$ \tilde{\A}([a],[c]).$ 
\end{tabbing}
As for any $a,b \in \A$, it is immediate that  
$\A(a,b) \leq \tilde{\A}([a],[b])$ then 
$\A$ defines a $\V$-category and the map
$a \mapsto [a]$ defines a $\V$-functor 
$\A \rightarrow \tilde{\A}$.
This map is surjective and defines actually a 
bisimulation from $\A$ to $\tilde{\A}$ as condition
$(1')$ is satisfied by the very definition
of $\tilde{\A}$.
\epf

We shall now relate bisimilarity to the existence 
of spans and cospans of surjective functionnal 
bisimulations. With cospans things are working 
well without extra assumptions.
\begin{prop}\label{cospan}
$\A$ and $\B$ are bisimilar if and only if there
exists a cospan of functionnal bisimulation
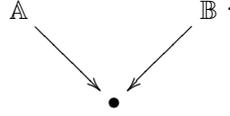
\xymatrix{ \A \ar[rd] & &  \ar[ld] \B\\
 & \bullet &}.
\end{prop}
\pf
We know already that such a span implies 
that $\A$ and $\B$ are bisimilar. It remains 
to prove the converse.
Suppose $\A$ and $\B$ bisimilar via $R \subseteq \A \times \B$.
If $\sim$ denotes the equivalence bisimulation
on $\A$ generated by $R^{-1} \circ R$ 
and also the equivalence on $\B$ generated 
by $R \circ R^{-1}$ then it happens that 
$\A/\sim \; \cong \; \B/\sim$.
To see this note that equivalence classes for $\sim$ on $\A$ 
correspond sequences $...a_i b_i a_{i+1} b_{i+1} ...$ 
where 
$(a_i, b_i) \in R$, $(a_{i+1},b_i) \in R$ and that contains
at least one $a_i$ and one $b_i$. Sequences as above
also correspond bijectively to equivalence classes
of $\sim$ on $\B$.
Let us write $\sim$ again for the bijection above between 
classes $[a] \in Obj(\A) /\sim$ and $[b] \in Obj(\B)/\sim$.
To see that the later
defines a $\V$-equivalence one needs to show:
$\A / \sim([a],[a']) = \B / \sim( [b], [b'] )$
for any $a,a',b,b'$ with $[a] \sim [b]$ and $[a'] \sim [b']$.
Given such $a,a',b,b'$, one may suppose $(a,b) \in R$ and 
$(a',b') \in R$ so that 
\begin{tabbing}
$(\A / \sim)([a],[a'])$\= $=$ \=$\bigvee_{a'' \sim a'} \A(a,a'')$\\
\>$\leq$ \>$\bigvee_{a'' \sim a'} \bigvee_{b' R a''} \B(b,b')$
$\;\;\;$ 
since $(a,b) \in R$\\
\>$\leq$ \>$\bigvee_{[b'] \sim [a']} \B(b,b')$ \\
\>$=$ \>$\B/ \sim ([b],[b'])$.
\end{tabbing}
\epf

To characterise the bisimilarity in terms of spans
of surjective functionnal bisimulations we considered
some extra assumptions on the base quantaloid. 
We shall call a quantaloid $\V$ 
{\em locally distributive} when the local preorders 
$\V(a,b)$ are distributive. 
\begin{prop}\label{span}If $\V$ is locally distributive,
$\V$-categories $\A$ and $\B$ are bisimilar 
if and only if there exists a span of functionnal bisimulation
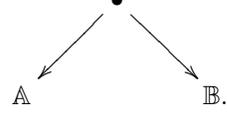
\xymatrix{  & \bullet \ar[ld] \ar[rd] &\\
\A & &  \B.}\\
\end{prop}
The claimed result follows from the following two lemmas.
\begin{lemma}
For any quantaloid $\V$, $\VCat$ has pullbacks.
\end{lemma}
\pf Given $\V$-categories $\A$ and $\B$, their (cartesian) product
$\A \times \B$ is as follows. Its set of objects is 
the subset cartesian product
$Obj(\A) \times Obj(\B)$ of pairs $(a,b)$ with $a_+ = b_+$, 
its homsets are given by the formula
$$( \A \times \B )((a,b),(a',b')) = \A(a,a') \wedge \B(b,b'),$$
the compositions
$$\mu_{ (a,b),(a',b') }: (\A \times \B) ((a,b), (a',b')) \otimes 
(\A \times \B) ((a',b'), (a'',b'')) \leq (\A \times \B)((a,b),(a'',b''))$$  
are given by:
\begin{tabbing}
\hspace{1.5cm}\=
\hspace{2cm}\=
$( \A(a,a') \wedge \B(b,b') ) \otimes ( \A(a',a'') \wedge \B(b',b'')
)$\\
\>$\leq$
\>$(\A(a,a') \otimes \A(a',a'')) \wedge ( \B(b,b') \otimes \B(b',b''))$\\
\>$\leq$
\>$( \A(a,a'') \wedge \B(b,b'') ).$
\end{tabbing}
and the units by:
$$I_{a_+} = I_{b_+} 
\leq (I_{a_+} \wedge I_{b_+}) 
\leq \A(a,a) \wedge \B(b,b).$$
Now given a diagram in $\VCat$, 
$$\xymatrix{ \A \ar[r]^F &  \C & \B \ar[l]_G},$$
its pullback is given by:
$$\xymatrix{ \A &   \A \wedge_\C B \ar[l]_{\bar{G}} \ar[r]^{\bar{F}} & \B }$$
where $\A \wedge_C \B$ is the subcategory of $\A \wedge \B$ 
generated by pairs $(a,b)$ with $F(a) = G(b)$ and the arrows
are the obvious embeddings.
\epf   
\begin{lemma}\label{dist}
If the quantaloid $\V$ is {\em locally distributive} then
pullback operation preserves the surjective functionnal
bismulation in $\VCat$.
\end{lemma}
\pf Consider the pullback diagram
\begin{center}
$
\xymatrix{ \A \wedge_{\C} \B \ar[r] \ar[d]_{\bar{G}} &  \B \ar[d]^G\\
\A \ar[r]_{F} & \C 
}
$
\end{center}
where $G$ is a surjective functionnal bisimulation.
Let $a \in \A$, since $G$ is surjective,
there exists $b \in \B$ such that 
$G(b) = F(a)$ and $\bar{G}(a,b) = a$. 
For such $a$ and $b$, given $a' \in \A$,
\begin{tabbing}$\hspace{2cm}$\= $\hspace{1cm}$\=
$\bigvee_{(a',b') \mid \bar{G}(a',b') = a'} 
(\A \wedge_{\C} \B)((a,b),(a',b'))$\\
\> $=$ \> 
$\bigvee_{(a',b') \mid F(a') = G(b')} \A(a,a') \wedge \B(b,b')$\\
\> $=$ \> $\A(a,a') \wedge \bigvee_{b' \mid F(a') = G(b')} \B(b,b')$
\;\;\;by distributivity\\
\> $=$ \> $\A(a,a') \wedge \C(Fa, Fa')$\\
\> $\geq$ \>$\A(a,a')$.
\end{tabbing}
\epf

Let $\Od$ denotes the sets of functional 
bisimulations
whose underlying maps on objects are surjections.
This class of maps forms a subcategory of $\VCat$.
It also satisfies a series of axioms
$(A1)$ to $(A6)$ which are the same or quite similar
to those stated by Joyal and Moerdijck \cite{JoMo94}. 
We shall review these axioms pointing out the differences 
with the axioms for open maps.\\

\begin{prop}(A1)
$\Od$ contains isomorphisms and is closed under 
composition.
\end{prop}
This is straightforward. Remember that $f:\A \rightarrow \B$ is 
an isomorphism in $\VCat$ 
if and only if its underlying map on objects is one-to-one
and $\forall a,a' \in \A, \A(a,a') = \B(f(a),f(a'))$.\\

Note also that $f$ is a split epi in $\VCat$ if and 
only if its underlying map on objects is surjective
and $\forall b \in \B$, one may find a $s(b) \in \A$ such that
$f(s(b)) = b$ and for all $b,b' \in \B$,  
$\B(b,b') = \A(s(b),s(b'))$.
Given a split epi $f: \A \rightarrow \B$ such that for each 
$a \in \A$, there exists a section $s$ for $f$ with value $a$
then $f$ belongs to $\Od$, as in this case for such an $s$
for any $b \in \B$,
$$\B(fa,b) = \A(s(f(a)),s(b)) = \A(a,s(b)).$$

As seen previously in \ref{dist},
\begin{prop}(A2 - stability axiom - preservation by pullback)\\
If $\V$ is {\em locally distributive},
in any pullback square
$$\xymatrix{  
\A \wedge_{\Cd} \B  \ar[r]^{\bar{f}} \ar[d]_{\bar{g}} & \B \ar[d]^g \\
\A \ar[r]_f & \Cd
 }$$ if $g$ belongs to
$\Od$ then $\bar{g}$ also belong to $\Od$.
\end{prop}

\begin{prop}(A3 - descent axiom)
In any pullback square
$$\xymatrix{  
\A \wedge_{\Cd} \B  \ar[r]^{\bar{f}} \ar[d]_{\bar{g}} & \B \ar[d]^g \\
\A \ar[r]_f & \Cd
 }$$ 
if $f$ is surjective on objects and satisfies the 
condition
\begin{center}
$(*)$ $\C(c,c') \leq \bigvee_{ f(a)=c, f(a')=c' } \A(a,a')$, for all $c,c'
\in \C$ 
\end{center}
and $\bar{g}$ 
belongs to $\Od$ then $g$ belongs to $\Od$.
\end{prop}
Note that any $f$ that is split epi satisfies the condition
$(*)$ above.
The descent axiom $(A3)$ in \cite{JoMo94} differs from the one above
on the point that $f$ is there just required $epi$.\\  
\pf
If $f$ satisfies $(*)$ 
then for any $b \in \B$ and $c \in \Cd$,
\begin{tabbing}
$\Cd(gb,c)$\= $\leq$ \=$\bigvee_{f(a)=g(b),f(a')=c'} \A(a ,a')$\\
\>$\leq$
\>$\bigvee_{a' \mid f(a') =c}
\bigvee_{b' \mid g(b') = c'} \A(a,a')  \wedge \B(b,b')$\\
\>$\leq$
\>$\bigvee_{b' \mid g(b') = c} \B(b,b')$
\end{tabbing}
\epf

$\VCat$ has a terminal object $1$ as follows. 
$1$ has one point $*_v$ per $v \in \V$ with $(*_v)_+ = v$
and homs given by $1(*_u, *_v) = \top_{u,v}$ the maximal element
of $\V(u,v)$.
Also for any small collection $(\A_i)_{i \in I}$ of enrichments
the coproduct $\coprod_{i \in I} \A_i$ exists, it has
set of objects the disjoint union $\coprod_{i \in I} Obj(\A_i)$
and its hom is given by the formula:
$(\coprod_{i \in I} \A_i)(x,y)$ $=$ $\A_i(x,y)$ if both 
$x$ and $y$ belongs to $\A_i$ or $\bot_{x_+,y_+}$ the least 
element of $\V(x_{+},y_{+})$ otherwise.

\begin{prop}(A4)
For any set $I$, the unique map 
$\coprod_{i \in I} 1 \rightarrow 1$ belongs to $\Od$.
\end{prop}
\pf
The coproduct $\coprod_{i \in I} 1$ has as objects
say the $*_{v,i}$'s where $v$ ranges in $\V$ and  $i$ in $I$
with $\coprod 1( *_{u,i}, *_{v,j} )$ is 
$\top_{u,v}: u \rightarrow v$ if $i=j$ and equals
$\bot_{u,v}$ otherwise.    
The unique map $!: \coprod_{i \in I} 1 \rightarrow 1$
sends any $*_{v,i}$ to the unique object $*_v \in 1$.
Given any $*_{u,i} \in \coprod 1$,
\begin{tabbing}  
$1(!(*_{u,i}),*_v)$ \=$=$ \=$\top_{u,v}$\\
\>$\leq$ \>$(\coprod 1)(*_{u,i},*_{v,i})$\\
\>$=$ 
\>$\bigvee_{x \in (\coprod 1) \mid !(x) = *_v} (\coprod 1)(*_{u,i},x)$.
\end{tabbing}
\epf

\begin{prop}
For any family of arrows $\A_i \rightarrow \B_i$ in $\Od$,
their sum $\coprod_{i \in I} \A_i \rightarrow \B_i$ belongs to
$\Od$.
\end{prop}
\pf
Consider a family $f_i : \A_i \rightarrow \B_i$ in $\Od$.
The sum $f = \coprod_{i \in I} f_i$ sends $x \in \A_i$ to 
$f_i(x) \in \B_i$.
Given $x \in \A_i$ and $y \in \B_j$,
we have to show
$$(\coprod_{i \in I} \B_i)\;(f_i(x),y)) \leq \bigvee_{x' \mid f(x') =
  y} (\coprod_{i \in I}\A_i)\;(x,x').$$
If $i \neq j$ then the left hand side term is $\bot$.
If $i = j$, then the left hand side term is 
$\B_i(f_i(x),y)$ that is less than $\bigvee_{x' \mid f_i(x') = y}
\A_i(x,x')$ as $f_i \in \Od$, that is less the right hand side term.  
\epf

\begin{prop}(A6-Quotient axiom)
In any commutative diagram
$$\xymatrix{ \A \ar[dr]_{g} \ar[rr]^{p} & & \B \ar[dl]^f \\
& \Cd & },$$
if $g \in \Od$ and $p$ is surjective then 
$f \in \Od.$\\ 
\end{prop}
\pf
Given a diagram as above with $g \in \Od$,
for any $b \in \B$ and $c \in \Cd$,
$\Cd(f(b),c) = \Cd(g(a), c)$ for some $a \in A$
with $p(a) = b$.
Then
\begin{tabbing} 
$\Cd(f(b),c)$
\=$\leq$ \=$\bigvee_{a' \mid g(a') = c} \A(a ,a')$\\
\>$\leq$ \>$\bigvee_{a' \mid g(a') = c} \B(b,p(a'))$\\
\>$\leq$ \>$\bigvee_{b' \mid f(b') = c} \B(b,b')$ 
\end{tabbing} 
\epf
\section{Change of Base}
The study of the change of base for enrichments over bicategories
gave rise to the concept of two-sided enrichments \cite{KLSS02}.
We will need some elements from of this theory (again in the 
particular and simpler case of enrichments over quantaloids). 
The main point here is the fact that all information about 
change-of-base is concealed
in the tricategory $Caten$ to be introduced below. We shall revisit
very quickly this simple tricategory admitting orders as 2-homs
before to state the change of base theorems. We refer the reader
to \cite{LaSc02} for a detailed presentation of this material.\\

Let $\spa$ be the bicategory of spans over $\set $ \cite{bena}.
Recall that the map $f:\, X\rightarrow Y$ in
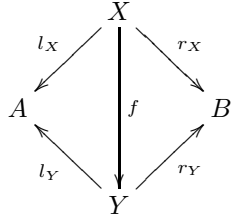
\begin{center} 
$\xymatrix{ &  X \ar[dd]^f \ar[dl]_{l_X} \ar[dr]^{r_X} & \\ 
A &         & B \\ &  Y \ar[ul]^{l_Y} \ar[ur]_{r_Y}}$ 
\end{center} is a 2-cell 
$f:\, \left(X,l_{X},r_{X}\right)\Rightarrow \left(Y,l_{Y},r_{Y}\right)$
in $\spa $ provided it makes everything in sight commute. A 2-cell
in $\spa $ is also called \emph{morphism of spans.} It is well
known that a span is a left
adjoint in $\spa $ precisely when its left leg is iso. It follows
that the iso class of a left adjoint in $\spa $ has a span with identity
left leg as a canonical representant. Right adjoints are characterized
similarly, i.e. their right legs are iso.

\subsection*{Caten }
$Caten$'s objects are locally ordered bicategories. 
An arrow $A:{\mathcal{V}}\rightarrow {\mathcal{W}}$
in $Caten$, called a \emph{two-sided enrichment}, consists of

\begin{itemize}
\item a span 
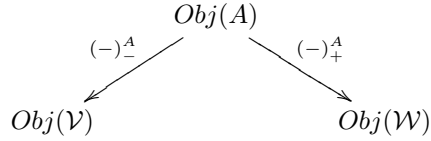
\begin{center} 
$\xymatrix{ & Obj(A) \ar[ld]_{ {(-)}_{-}^A } \ar[rd]^{ {(-)}_{+}^A } &
  \\ 
Obj( {\cal V} ) &    &  Obj( {\cal W} )  }$ 
\end{center}
in $Set$. 
Let $\widehat{A}\deq \left(Obj\left(A\right),
\left(-\right)_{-}^{A},\left(-\right)_{+}^{A}\right)$;
\item for all $a,b\in Obj\left(A\right)$ a collection of functors (monotone
maps) \[
A_{a,b}:{\mathcal{V}}(a_{-},b_{-})\rightarrow {\mathcal{W}}(a_{+},b_{+})\]
 such that
\end{itemize}
\begin{center}  
$\xymatrix{
{\cal V}(a_{-}, b_{-}) \times {\cal V}(b_{-}, c_{-})
\ar[d]_{A_{a,b} \times A_{b,c}}
\ar[r]^<(.3){\otimes^{\cal V} }  
\ar@{}[rd]|{\leq}
& {\cal V}(a_{-}, c_{-})
\ar[d]^{A_{a,c}}
\\
{\cal W}(a_{+}, b_{+}) \times {\cal W}(b_{+}, c_{+})
\ar[r]_<(.3){\otimes^{\cal W} }  
& {\cal W}(a_{+}, c_{+})
}$
\end{center} for any $a,b,c\in A$
and such that \begin{center} 
$\xymatrix{ 
1 \ar@{=}[d] \ar[rr]^{ id_{a_{-}} } 
\ar@{}[rrd]|{\leq} && 
{\cal V}(a_{-}, a_{-}) 
\ar[d]^{A_{a,a}}  
\\ 
1  
\ar[rr]_{ id_{a_{+}} }  
& & {\cal W}(a_{+}, a_{+}) 
}$
 \end{center} for any
$a\in A$.\\

For any two-sided enrichments ${\mathcal{V}}$ and ${\mathcal{W}}$,
the bicategory $Caten({\mathcal{V}},{\mathcal{W}})$ is locally partially
ordered. Let $A,B\in Caten({\mathcal{V}},{\mathcal{W}})$. A 2-cell
$f:\, A\Rightarrow B$ in $Caten$ consists of a morphism of spans
$f:\, \widehat{A}\rightarrow \widehat{B}$ such that

\[
A_{a,b}\leq B_{fa,fb}:{\mathcal{V}}(a_{-},b_{-})\rightarrow {\mathcal{W}}(a_{+},b_{+})\]
 for any $a,b\in A$.\\ 

\noindent Vertical composition of 2-cells in $Caten$ is determined by $\spa $'s
one. 2-cells in $Caten$ are ordered as follows: 
$$f \leq g:A\Rightarrow B:{\mathcal{V}}\rightarrow {\mathcal{W}}$$
when \[
id_{a_{+}}\leq B_{fa,ga}\circ id_{a_{-}}:a_{+}\rightarrow a_{+}\]
 for any $a\in A$. 

Let $A:{\mathcal{U}}\rightarrow {\mathcal{V}}$ and $B:{\mathcal{V}}\rightarrow {\mathcal{W}}$
be two-sided enrichments. Their composite $B\circ A:{\mathcal{U}}\rightarrow {\mathcal{W}}$
is given by

\[
\widehat{B\circ A}\deq \widehat{B}\circ \widehat{A}\]
and by the functors ${B\circ A}_{(a,b),(a',b')}$ being the composites
\[
{\mathcal{U}}(a_{-},{a'}_{-})\xrightarrow{{A_{(a,a')}}}{\mathcal{V}}(a_{+},{a'}_{-})={\mathcal{V}}(b_{-},{b'}_{-})\xrightarrow{{B_{(b,b')}}}{\mathcal{W}}(b_{+},{b'}_{-})\]

Horizontal composition of 2-cells in $Caten$ is determined by $\spa $'s
one.

\subsection*{Adjoints in Caten}
Amongst the nice properties of $Caten$ - proved in \cite{KLSS02},
(and detailed in \cite{LaSc02} for locally preordered bases) - let us 
mention that the cartesian product of locally partially ordered 
bicategories extends straightforwardly to a pseudo functor 
$Caten\times Caten\rightarrow Caten$
that makes $Caten$ into a monoidal tricategory. With that 
monoidal structure $Caten$ is biclosed.
More importantly for our concern,
the change of base 
bicategory, one has the following representability
property:
\begin{prop}
For any quantaloid ${\mathcal{V}}$, the 2-category $\VCat $ is representable: \[
\VCat \cong Caten(1,{\mathcal{V}})\]
\end{prop}

As expected, adjointness in $Caten$ is by and large determined by
what happens in $\spa $. A two-sided enrichment $F:\, \V \rightarrow \W $
is a left adjoint provided
\begin{enumerate}
\item $F$ is a 
2-functor $\V \rightarrow \W$
(so in particular $Obj\left(A\right)=Obj\left(\V \right)$);
\item the functors $F_{u,v}:{\mathcal{V}}(u,v)\rightarrow {\mathcal{W}}(Fu,Fv)$
have {\em right adjoints} $G_{u,v}:{\mathcal{W}}(Fu,Fv)\rightarrow {\mathcal{V}}(u,v)$.
\end{enumerate}
One may check that a for a lax functor $F: \V \rightarrow \W$
as above that satisfies condition $2.$ then $F$ satisfies 
condition $1.$ if and only if the collection of
local adjoints $G_{u,v}$ satisfies the following two conditions:
\begin{enumerate}
\item $G_{u,v}(f) \otimes G_{v,t}(g) 
\leq G_{u,t}(f \otimes g)$ for all $u,v,t\in {\mathcal{V}}$
for all $f:Fu\rightarrow Fv$ and $g:Fv\rightarrow Ft$ ;
\item $id_{u}\leq G_{u,u}\circ id_{Fu}$ for all $u$ of ${\mathcal{V}}$.
\end{enumerate}

$\;$\\
Using the representability for ${\mathcal{V}}$-categories and the
fact that $Caten$ is a tricategory, one obtains a change of base theorem
for enrichments over bicategories. Postcomposition with any two-sided
enrichment $F:{\mathcal{V}}\rightarrow {\mathcal{W}}$ yields a 2-functor
$F\circ -:Caten(1,{\mathcal{V}})\rightarrow Caten(1,{\mathcal{W}})$,
and by the representability result a 2-functor $F_{@}:\VCat \rightarrow \WCat $
given by:

\begin{itemize}
\item $Obj(F_{@}\A)=\{(a,x)\mid a\in Obj(\A),x\in Obj(F),a_{+}=x_{-}\}$;
\item ${(a,x)}_{+}^{F_{@}\A}=x_{+}$;
\item $F_{@}\A((a,x),(b,y))=F_{x,y}(\A(a,b))$.
\end{itemize}
For any ${\mathcal{V}}$-functor $f:\A\rightarrow \B$, $F_{@}(f)$
is the ${\mathcal{W}}$-functor with underlying map $(a,x)\mapsto (fa,x)$.

In the same way, any 2-cell $\sigma :F\rightarrow G$ of $Caten$
yields a 2-natural transformation $\sigma _{@}:F_{@}\rightarrow G_{@}:\VCat \rightarrow \WCat $. 

\begin{thm}
There is a pseudo-functor $(-)_{@}:\, Caten\rightarrow \TwoCat $
that sends a locally ordered bicategory ${\mathcal{V}}$ to the 2-category
$\VCat $. Its actions on two-sided enrichments and 2-cells are the
ones described above. 
\end{thm}
Of particular interest to this investigation, one gets an adjoint pair
$F_{@}\dashv G_{@}$ in $\TwoCat $ from an adjoint pair $F\dashv G$
in $Caten$. (Recall from the previous section that a left adjoint 
in $Caten$ is given by a peculiar collection of 
local left adjoints)

\begin{thm}
\label{adjVWCat} Let $\V $ and $\W $ be quantaloids and $F:{\mathcal{V}}\rightarrow {\mathcal{W}}$
a {\em 2-functor with local adjunctions} $F_{u,v}\dashv G_{u,v}$ for
all $u,v\in Obj({\mathcal{V}})$. Then the 2-functor $F_{@}$ admits
a right 2-adjoint, denoted $F^{@}$ and given on objects by 
\end{thm}
\begin{itemize}
\item $Obj(F^{@}\B)=\{(b,v)\mid b\in Obj(\B),v\in Obj({\mathcal{V}}),b_{+}=Fv\}$;
\item ${(b,v)}_{+}=v$;
\item $F^{@}\B((b,v),(b',v'))=G_{v,v'}(\B(b,b'))$.\\

\end{itemize}

{\bf Example 1}\\
Betti's automata will provide our first examples of
change of base.
Let $\M$ and $\N$ be monoids. A {\em congruence} relation 
$r:\M \nrightarrow \N$ verifies  \begin{tabbing}
- \(\forall (m,n), (m',n') \in r, 
(m \cdot m' , n \cdot n') \in r\),\\
- \((id_{\M}, id_{\N}) \in r.\)
\end{tabbing} Note that congruences generalize the usual monoid
morphisms $\M \rightarrow \N$.
For any such $r$, there is a monoidal functor 
$C(r):C(\M)\rightarrow C(\N)$ taking $
L\subseteq \M$ to the direct image 
$\{n\, \mid \, (m,n)\in r\, \wedge \, m\in L\}$ of $L$ by $r$. 

An important point is that the functor $C(r)$ has a right adjoint,
namely $R(r)$ defined for $K\subseteq \N$ by \[
R(r)(K)=\{m\in \M\, \mid \, \forall n\in \N,\, (m,n)\in r\Rightarrow n\in K\}\]
This right adjoint fails to be monoidal for a general $r$. 

Consider now a monoid morphism $f:\M \rightarrow \N$. Then for the relation
$r\subseteq \N \times \M$ given by $(n,m)\in r\Leftrightarrow f(m)=n$,
$C(r)$ corresponds to the inverse image functor $f^{-1}:C(\N)\rightarrow C(\M)$.
So this functor has a right adjoint traditionally written $\forall _{f}$
and given by \[
\forall _{f}(L)=\{n\mid \forall m\in \M,(f(m)=n\Rightarrow m\in L)\}\]
for all $L\subseteq \M$. In case when the relation $r\subseteq \M\times \N$
is $(m,n)\in r\Leftrightarrow n=f(m)$, $C(r)$ corresponds to the
left adjoint $\exists _{f}$ of $f^{-1}:C(\N)\rightarrow C(\M)$. One
may check that $\exists _{f}$ is \emph{strong} (i.e. it preserves
strictly the monoidal structure) and thus is a left adjoint in $Caten$.
Then $\left(\exists _{f}\right)_{@}$ is a left 2-adjoint by theorem
\ref{adjVWCat}.\\

{\bf Example 2}\\
The previous example with monoids suggests an immediate
generalization to categories (monoids with many points!).
Define a congruence $r$ between categories
$\C$ and $\Db$ - still denoted by $r: \C \nrightarrow \Db$ -
as a  span
\begin{center}
$\xymatrix{ & r_0 \ar[ld]_{(-)_{-}} \ar[rd]^{(-)_{+}} &\\
Ob(\C) &     & Ob(\Db) }$
\end{center}
and a collection of relations 
$r_{x,y} \subseteq \C(x_{-},y_{-}) \times
\Db(x_{-},y_{-})$, $x$,$y$ ranging in $r_0$ such that:
\begin{itemize}
\item $(id_{x_-},id_{x^+}) \in r_{x,x}\;for\;all\;x \in r_0$;     
\item if $(f,f') \in r_{x,y}$ and $(g,g') \in r_{y,z}$
then $(g \circ f, g' \circ f') \in r_{x,z}$,
\;\;for\;all\;$x,y,z\in r_0$ and all 
$f:x_{-} \rightarrow y_{-}$,  
$f':x_{+} \rightarrow y_{+}$,
$g:y_{-} \rightarrow z_{-}$ and 
$g':y_{+} \rightarrow z_{+}$.
\end{itemize}

One obtains a two-sided enrichment $B(r): B(\C) \rightarrow B(\Db)$
with $Obj(B(r)) = r_0$ and for objects $x,y$ of $r$, for  
any $L \subseteq \C(x_{-}, y_{-})$, 
$$B(r)_{x,y}(L) = \{g \mid (f,g) \in r_{x,y} \subseteq \Db(x_+,y_+).$$
Each $B(r)_{x,y}: B(\C)(x_{-},y_{-}) \rightarrow B(\Db)(x_{+},y_{+})$, 
as defined above, admits a right adjoint namely $R(r)_{x,y}$ sending
$K \subseteq \B(x_{+},y_{+})$ to 
$$\{l: x_{-} \rightarrow y_{-} \in \C
\mid \forall k:x_{+} \rightarrow y_{+}, (l,k) \in r_{x,y} \Rightarrow
k \in K \} \subseteq \C(x_{-},y_{-}).$$
Nevertheless the data $R(r)_{x,y}$ for $x,y$ ranging in $r_0$ fails to 
define a two-sided enrichements in general.\\
  
Consider now a functor $f: \C \rightarrow \Db$.
Then for the congruence
$\Db \rightarrow \C$
defined by the relations 
$r_{x,y} \subseteq \Db(f(x),f(y)) \times \C(x,y)$ 
defined by $(k,l) \in r_{x,y}$ if and only if
$f(l) = k$, yields a two sided enrichement
$B(r): B(\Db) \rightarrow B(\C)$ that we shall denote 
again $f^{-1}$.  It has local right adjoints
$R(r)_{x,y}$ still denoted $\forall_{x,y}$
sending $L \subseteq \C(x,y)$
to $$\{ k: f(x) \rightarrow f(y) \mid \forall l: x \rightarrow y,
f(l) = k \Rightarrow l \in L \}.$$
In case when $r$ denotes the inverse congruence $\C \rightarrow \Db$,
$B(r): B(\C) \rightarrow B(\Db)$ is a 2-functor that we shall 
denote $\exists_f$, it has a local right adjoints 
given by the $f^{-1}_{x,y}$ above, so this is to say 
that $\exists_f \dashv f^{-1}$ in $Caten$.\\  

{\bf Example 3}\\
We shall come back now to the isomorphim \ref{comma}.
For any $\V$-functor $f: \A \rightarrow \B$, 
the functor $\V(f): \V(\A) \rightarrow \V(\B)$ is actually a left 
adjoint in $Caten$. This is easy to see. The left adjoints 
to the $\V(f)_{a,b}: \V(a_+,b_+) \downarrow \A(a,b) \rightarrow 
\V(a_+,b_+) \downarrow \B(fa,fb)$ are the $G_{a,b}$
sending $y \leq \B(fa,fb)$ to $min \{ y, \A(a,b) \} \leq \A(a,b)$.
Now for any $a,b,c \in \A$, 
\begin{tabbing}
$G_{a,b}(x) \otimes G_{b,c}(y)$\=$=$ 
\=$min \{x, \A(a,b) \} \otimes min \{y, \A(b,c) \}$\\
\>$\leq$
\>$min \{ x \otimes y, \A(a,b) \otimes \A(b,c) \}$\\
\>$\leq$ $min \{x \otimes y, \A(a,c) \}$\\
\>$=$ $G_{a,c} ( x \otimes y )$
\end{tabbing}
which is the coherence condition 1. for the local right 
adjoints. 
The coherence condition 2. amounts to 
$id_{a_+} \leq min \{ id_{a_+} , \A(a,a) \}$ for all
$a \in \A$, which clearly holds.\\

In the above  situation the change of base $\V(f)_@ : \V(\A)-Cat \rightarrow 
\V(\B)-Cat$ corresponds exactly via the isomorphism \ref{comma}
to the functor 
$\VCat \downarrow \A  \rightarrow \VCat \downarrow \B$ given by
composition with $f$, whereas the adjoint $\V(f)^@$ corresponds (via 
\ref{comma}) to the pullback along $f$ functor 
$\VCat \downarrow \B  \rightarrow \VCat \downarrow \A$.

\section{Bisimulations and change of base}
It seems natural to consider simulations/bisimulations
up to change of base and to ask when the changes of base
preserves/reflects bisimularity. 
We shall give a simple criterion for the preservation to
happen.
\begin{prop}
\label{pres} Let $\V $ and $\W $ be quantaloids. 
Any 2-sided enrichment $F:\V \rightarrow \W $
with local right adjoints induces a change
of base $F_@$ that preserves the class $\Od$ (and thus bimularity). 
\end{prop}
\pf Consider an open $f: \A \rightarrow \B$ in $\VCat$.
If the underlying map of $f$ is surjective then 
also is the underlying map of $F_@(f)$. 
Then for any $(a,x) \in F_@(\A)$ and $(b,y) \in F_@(\B)$
 \begin{tabbing}
$F_@(\B)( F_@(f)(a,x),(b,y) )$ \=$=$ \=$F_{x,y}( \B(fa,b) )$  \\
\>$=$ \>$F_{x,y}(  \bigvee_{ \{ a' \in A \mid f(a') = b \} } \A(a,a') )$\\
\>$=$ \>$\bigvee_{ \{ a' \in \A \mid f(a') = b \} } F_{x,y}(\A(a,a'))$
$\;\;\;$since $F_{x,y}$ is left adjoint\\
\>$=$ \>$\bigvee_{ \{ (a',y') \in F_@(\A) \mid F_{@}(f)(a',y') = (b,y) \} } 
F_@(\A)((a,x),(a',y'))$.
\end{tabbing}
\epf
 
As a consequence of this any left adjoint $F$ in $Caten$
will induce a change of base $F_@$ that preserves bisimularity.
Our previous examples provides change of base preserving 
bisimilarity. First if $f: \C \rightarrow \Db$ is a functor 
then $\exists_f$ (that is left adjoint) but also $f^{-1}$,
that has local right adjoints, will both induce change of bases
preserving bismularity.
For any $\V$-functor $f:\A \rightarrow \B$, the $\V$-functor
$\V(f): \V(\A) \rightarrow \V(\B)$ being left adjoint
${\V(f)}_@$ preserves bisimilarity. Which is no suprise here
as the bisimularity in $\V(\A)-Cat$ correspond to the 
bisimularity over $\A$ and its perservation by $\V(f)$
is equivalent to the fact that any bisimilar arrows $h,k$ 
over $\A$ yields bisimilar arrows $h \circ f, k \circ f$ over $\B$.    

\section{Refinement of Specifications }

In this section, we elaborate on an extended example illustrating
the use of the categorical machinery introduced so far. We advocate
that enriched categories are a convenient framework for the deployment
of the so-called {\em categorical transition systems} \cite{kw:ctcs}\cite{kw:cmcim}\cite{koslo},
in the sense that \emph{coherence conditions} are taken care of by
the enriched structure. We then apply the rest of the machinery to
study refinements of specifications in this framework. 

\begin{defn}
\label{def-superlax}Let $\ka $ be a bicategory. The category $\slam {\ka }$
is given by the data
\begin{enumerate}
\item Objects: normalized pseudo-functors from a free categories $\free
  G$ over graphs $G$ to $\ka $;
\item Morphisms: given $s:\, \free G\rightarrow \ka $ and $t:\, \free H\rightarrow \ka $
normal pseudo-functors, a morphism $\alpha :\, p\rightarrow q$
is given by a graph morphism $k:\, G\rightarrow H$ and a lax transformation
$\alpha :\, s\Rightarrow t\circ \free k$ with left adjoint components;
\item Composition: $\beta \circ \alpha =\left(l\circ k,l\beta \circ \alpha \right)$
where $\alpha :\, s\Rightarrow t\circ \free k$ and $\beta :\, t\Rightarrow u\circ \free l$
while the vertical composition $l\beta \circ \alpha $ is given by
componentwise pasting.
\end{enumerate}
\end{defn}
Let $\T $ be a category with finite limits and $\spn {\T }$
bicategory of spans over $\T $ \cite{bena}. We call \emph{categorical
transition systems} or \emph{cts}'s \emph{over} $\T $ objects of
$\slam {\spn {\T }}$. They are essentially generalized labelled transition
systems where the labels are organized in $\spn {\T }$. As an example,
consider the imperative program in-context

\begin{lyxcode}
x:=20;while~x~>~0~do~x:=~x-1~end~$\left[x:nat\right]$
\end{lyxcode}
\noindent It gives rise to a pseudo-functor $p:\, \free G\rightarrow \spn {\set }$
generated as follows \begin{center}

$
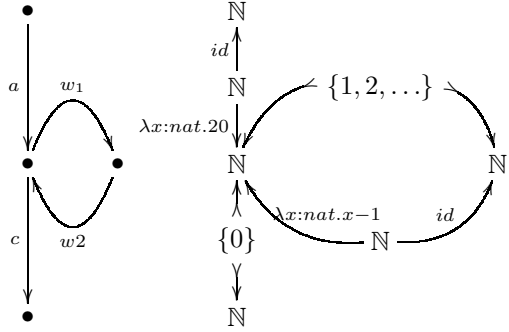
\xymatrix@=3mm{
\bullet \ar[dd]_{a} &  &&& \N &&&
\\
&&&& \N \ar[u]^{\one} \ar[d]_{\lambda x:nat.20} &
**{++} \{1,2,\ldots\}
\ar@/_1.2pc/@{>->}[dl] \ar@/^1.2pc/@{>->}[dr] &&
\\
\bullet \ar[dd]_{c} \ar@/^2pc/[rr]^{w_1} & &
\bullet \ar@/^2pc/[ll]^{w2} & & \N & & \N &
\\
&&&& **{++} \{0\} \ar@{>->}[u] \ar@{>->}[d] & 
\N \ar@/^1pc/[ul]_{\lambda x:nat.x-1} \ar@/_1pc/[ur]^{\one} &&
\\
\bullet & && & \N &&&
}$

\end{center} 

We see at hand of this example that the category $\T $ plays the
r\^ole of the type theory underlying the computation performed by
a cts. The states of a cts are labelled by $\T $'s objects i.e. types.
These types are those of the variables in scope. The legs of the spans
are labelled by terms, jointly representing a generalized transition
relation. Cts's allow a quite realistic modelling of imperative programs
including communication over typed channels. Moreover, the view of
a cts over $\T $ as an object of $\slam {\spn {\T }}$ offers a compact
representation of such programs. This fact was for instance expoited
in the design and the implementation of a deductive modelchecker 
\cite{memo-book}
(indeed, the representations in question could be accomodated by the
theorem prover PVS acting as a {}``logical back-end''). However,
one has to cope with coherence conditions when it comes down to calculations.

We shall now propose an alternative view of a cts. It is 
particularly interesting when addressing the question 
of refinement of specifications that we shall interpret as
a functor between categories of types $\T \rightarrow \T'$.\\ 

If $M \subseteq \C_{0}$
is a collection of objects of a category $\C$ we write $M\downarrow $
the sieve (\emph{crible},
right-ideal) generated by $M$.
\begin{defn}
Let $\T $ be a category with finite limits. The quantaloid $S\left(\T \right)$
is given by
\begin{itemize}
\item objects: $\ob {\T }$;
\item $S\left(\T \right)\left(X,Y\right)$ is 
the set of generated cribles $M \downarrow$
for $M \subseteq \spn {X,Y}$, ordered by inclusion;
\item the horizontal composition is given by the formula
$$M \downarrow \circ N \downarrow = (M \circ N) \downarrow$$
where $\circ$ is the pointwise composition of set of spans. 
\end{itemize}
\end{defn}
A cts $p:\, \free G\rightarrow \spn {\T }$ over $\T $ gives rise
to an $S\left(\T \right)$-category $\A _{p}$ as follows:

\begin{itemize}
\item $\ob {\A _{p}}$ is the set of $p$'s states, i.e. $G_{0}$;
\item $\A _{p}\left(x,y\right)$ is 
$\left\{ p\left(f\right)\, \mid \, f\in \free
  G\left(x,y\right)\right\} 
\downarrow$.
\end{itemize}

Now given a {\em right exact} functor $F: \A \rightarrow \B$
between finitely complete categories, one defines an adjoint
$S(F): S(\A) \rightarrow S(\B)$ in Caten as follows.
$S(F)$'s action is given on objects by 
\textbf{$S(F)(x)\deq F(x)$}, and
on arrows by\[
\begin{array}{rlcl}
 S(F)_{x,y}: & S(\A)(x,y) & \rightarrow  & S(\B )(Fx,Fy)\\
  & M & \mapsto  & \{ (Fa,F(d_{a}),F(c_{a}) )\, \mid \, 
(a,d_{a},c_{a})\in M \} \downarrow \end{array}
\]
It is easy to see that $F$ is a 2-functor.
For each $x,y \in \A $, there is a ``local'' adjunction 
$S(F)_{x,y}\dashv U_{x,y}$
with right adjoints given by\[
\begin{array}{rlcl}
 U_{x,y}: & S(\B)(F(x),F(y)) & \rightarrow  
& S(\A)(x,y)\\
  & N & \mapsto  & \{ (a,d_a,c_a)  \mid 
  (F(a),F(d_a),F(c_a)) \in N    \}  \end{array}
\]

Moreover in the case of an adjunction
$F \dashv G: \A \rightarrow \B$,
$S(G)$ (that is already left adjoint in $Caten$) 
also admits {\em local left adjoints}. To see this let
$\overline{t}:F(a)\rightarrow b$ denote the transposed 
of $t:\, a \rightarrow G(b)$ across the adjunction, then
for any $x,y \in \B$,
the left adjoint to $S(G)_{x,y}$ is given by:
\[
\begin{array}{rlcl}
 R_{x,y}: & S(\A)( G(x),G(y) ) & \rightarrow  
& S(\B)(x,y)\\
  & N & \mapsto  & \{ (F(a),\overline{c_a},\overline{d_a})\, \mid \,
  (a,c_a,d_a)\in N \} \downarrow \end{array}
\]

According to the result above (bi)similar specifications
of categorical transition systems will have (bi)similar refinements
provided the refinement functor is {\em left exact}.

\section{Concluding Remarks }

We hope that the present work exhibits some pertinence of
the interaction enriched category theory and simulation/bisimulation 
theory. Extending Betti's work, we have shown that 
bicategorical enrichments over quantaloids can accomodate a wide 
spectrum of existing notions of automata and communicating processes 
including labelled transition systems, themselves
a special case of Betti's automata, and also categorical transition
systems. We then presented a few new results about two-sided enrichments
and exhibited applications measuring the impediments to the existence
of good change-of-base homomorphims. After having introduced 
an appropriate
notion of (bi)simulation for enrichemnts, {}``good'' turned out to mean {}
``bisimularity-preserving''. We illustrated the notion with an
example about refinements of specifications in the framework 
of categorical transitions systems. 

All the material is indeed quite formal and it is precisely the whole
point of the paper that it \emph{should} be formal. In other words,
our wish is that category theory is revelatory for structural properties.

This investigation represents only the beginning of a research with
generalized automata and their properties as subject. We expect the
framework flexible enough to accomodate more elaborated programming
constructs. We also intend to introduce a notion of \emph{homotopy
of paths} within a generalized automaton.
In a different perspective, we plan to study the notion of
bisimulation itself in terms of homotopies \cite{getco}.

\end{document}